\documentclass[12pt]{article}
\usepackage{graphicx,amssymb,amsfonts,amsmath,epsfig,color}

\makeatletter
\DeclareSymbolFont{AMSb}{U}{msb}{m}{n}
\DeclareSymbolFontAlphabet{\mathbb}{AMSb}
\renewcommand{\section}{\@startsection{section}{1}{\z@}%
                                    {-7ex \@plus -1ex \@minus -.2ex}%
                                    {2.5ex \@plus.2ex}%
                                    {\normalfont\large\scshape\centering}}
\renewcommand{\subsection}{\@startsection{subsection}{2}{\z@}%
                                       {-5ex \@plus -1ex \@minus -.2ex}%
                                       {1.5ex \@plus.2ex}%
                                       {\normalfont\normalsize\scshape}}
\renewcommand{\subsubsection}{\@startsection{subsubsection}{3}{\z@}%
                                       {-5ex \@plus -1ex \@minus -.2ex}%
                                       {1.5ex \@plus.2ex}%
                                       {\normalfont\normalsize\scshape}}

\renewcommand\@seccntformat[1]{\ignorespaces\csname #1name\endcsname\space
                               \csname the#1\endcsname.\quad}   
\newdimen\captionmargin
\newdimen\captionindent
\newdimen\captionwidth
\newcommand{\captionfont}{\slshape}
\newcommand\@captionlabel[1]{\textsc{#1:}\space}
\long\def\@makecaption#1#2{%
  \vskip\abovecaptionskip
  \captionwidth\hsize
  \advance\captionwidth -2\captionmargin
  \sbox\@tempboxa{\@captionlabel{#1}\captionfont #2}%
  \ifdim \wd\@tempboxa >\captionwidth
    \ifdim\captionindent>\z@
      \advance\captionwidth -\captionindent
      \hskip\captionindent
    \fi
    \hskip\captionmargin
    \parbox[t]{\captionwidth}{\leavevmode\hskip-\captionindent
      \@captionlabel{#1}\captionfont #2}%
  \else
    \global \@minipagefalse
    \hb@xt@\hsize{\hfil\box\@tempboxa\hfil}%
  \fi
  \vskip\belowcaptionskip}
\def\eqnarray{%
   \stepcounter{equation}%
   \def\@currentlabel{\p@equation\theequation}%
   \global\@eqnswtrue
   \m@th
   \global\@eqcnt\z@
   \tabskip\@centering
   \let\\\@eqncr
   $$\everycr{}\halign to\displaywidth\bgroup
       \hskip\@centering$\displaystyle\tabskip\z@skip{##}$\@eqnsel
      &\global\@eqcnt\@ne$\;\hfil{##}$\hfil
      &\global\@eqcnt\tw@$\;\displaystyle{##}$\hfil\tabskip\@centering
      &\global\@eqcnt\thr@@ \hb@xt@\z@\bgroup\hss##\egroup
         \tabskip\z@skip
      \cr}
\ifcase \@ptsize
\or
\or
\fi
  \divide\@tempdima\baselineskip
  \@tempcnta=\@tempdima
\makeatother
\begin{document}

\renewcommand{\theequation}{\arabic{section}.\arabic{equation}}
\renewcommand{\thefigure}{\arabic{figure}}
\newcommand{\gapprox}{%
\mathrel{%
\setbox0=\hbox{$>$}\raise0.6ex\copy0\kern-\wd0\lower0.65ex\hbox{$\sim$}}}
\textwidth 165mm \textheight 220mm \topmargin 0pt \oddsidemargin 2mm
\def\ib{{\bar \imath}}
\def\jb{{\bar \jmath}}

\newcommand{\ft}[2]{{\textstyle\frac{#1}{#2}}}
\newcommand{\be}{\begin{equation}}
\newcommand{\ee}{\end{equation}}
\newcommand{\bea}{\begin{eqnarray}}
\newcommand{\eea}{\end{eqnarray}}
\newcommand{\Identity}{{1\!\rm l}}
\newcommand{\cx}{\overset{\circ}{x}_2}
\def\CN{$\mathcal{N}$}
\def\CH{$\mathcal{H}$}
\def\hg{\hat{g}}
\newcommand{\bref}[1]{(\ref{#1})}
\def\espai{\;\;\;\;\;\;}
\def\zespai{\;\;\;\;}
\def\avall{\vspace{0.5cm}}
\newtheorem{theorem}{Theorem}
\newtheorem{acknowledgement}{Acknowledgment}
\newtheorem{algorithm}{Algorithm}
\newtheorem{axiom}{Axiom}
\newtheorem{case}{Case}
\newtheorem{claim}{Claim}
\newtheorem{conclusion}{Conclusion}
\newtheorem{condition}{Condition}
\newtheorem{conjecture}{Conjecture}
\newtheorem{corollary}{Corollary}
\newtheorem{criterion}{Criterion}
\newtheorem{defi}{Definition}
\newtheorem{example}{Example}
\newtheorem{exercise}{Exercise}
\newtheorem{lemma}{Lemma}
\newtheorem{notation}{Notation}
\newtheorem{problem}{Problem}
\newtheorem{prop}{Proposition}
\newtheorem{rem}{{\it Remark}}
\newtheorem{solution}{Solution}
\newtheorem{summary}{Summary}
\numberwithin{equation}{section}
\newenvironment{pf}[1][Proof]{\noindent{\it {#1.}} }{\ \rule{0.5em}{0.5em}}
\newenvironment{ex}[1][Example]{\noindent{\it {#1.}}}

\thispagestyle{empty}


\begin{center}

{\LARGE\scshape Singularity-free gravitational collapse and asymptotic safety
\par}
\vskip15mm

\textsc{Ram\'{o}n Torres\footnote{E-mail: ramon.torres-herrera@upc.edu}
}
\par\bigskip
{\em
Department of Applied Physics, UPC, Barcelona, Spain.}\\[.1cm]

\vspace{5mm}

\end{center}

\begin{abstract}
A general class of quantum improved stellar models with interiors composed of non-interacting (\textit{dust}) particles is obtained and analyzed in a framework compatible with asymptotic safety. First, the effective exterior, based on the Quantum Einstein Gravity approach to asymptotic safety is presented and, second, its effective compatible dust interiors are deduced. The resulting stellar models appear to be devoid of shell-focusing singularities.
\end{abstract}




\setcounter{equation}{0}

\section{Introduction}

Some time ago, it was suggested by Steven Weinberg \cite{Wein} that a quantum theory of gravitation may dynamically evade the divergences found in perturbative gravity. Specifically, this scenario, called \textit{asymptotic safety}, implies the UV completion of gravity based on a non-Gaussian fixed point of the Renormalization Group flow. At the present time and thanks to the advent of new functional renormalization group methods, there is accumulating evidence in favor of the asymptotic safety scenario (see \cite{N&R}\cite{R&S} and references therein); however, there are still some aspects of the approach that need clarification.

One of these aspects is that, since it seems only natural to demand that a truly fundamental theory of quantum gravity should be devoid of singularities, asymptotic safety should be able to provide singularity-free solutions as the result, for example, of a gravitational collapse.
However, it is still a mystery whether and how the singularities that appeared in General Relativity (GR) would be avoided in the asymptotic safety scenario. The difficulty relies on the complexity of a full approach to the collapse of matter in the framework of asymptotic safety. In fact, to my knowledge, there has only been one previous approximation to this problem \cite{Casadio} which suggests that the deviations from GR offered by the asymptotic safety approach could be too small to prevent the generation of singularities during gravitational collapse.

This work aims to contribute to the analysis of the presence/absence of singularities in the framework of asymptotic safety. In particular, an attempt will be made to obtain and analyze the general class of stellar models consisting of non-interacting particles which are compatible with asymptotic safety.
It is worth recalling that, in the framework of GR, the class of spherically symmetric solutions consisting of non-interacting (or \emph{dust}) particles are known as the Lemaitre-Tolman-Bondi (LTB) solutions and have been thoroughly studied (see, for example, \cite{G&P} and references therein). Since there is nothing preventing the collapse in these classical models, once the particles start collapsing they will be eventually forced to generate a singularity. From this classical point of view, the only question is whether this singularity will be space-like and hidden from any observers (as in the Oppenheimer-Snyder model \cite{O&S}) or it will form a naked singularity visible to, at least, some observers. In fact, it has been shown \cite{E&S}\cite{Christ} that the class of the LTB models is wide enough to admit both hidden and (locally or globally) naked singularities.
The final goal of this work is to show that the dust models compatible with asymptotic safety, unlike their analogous classical LTB models, are singularity-free.

The work has been divided as follows. In section \ref{secISS} the improved stellar exterior coming from the asymptotic safe approach is presented. Then, in section \ref{secInterior} the general class of dust interiors compatible with this exterior and with asymptotic safety is deduced. These solutions are analyzed in section \ref{secAbsSing} in search of matter or curvature singularities. Finally, the results are discussed in the concluding section \ref{secConc}.

\section{Exterior: Improved Schwarzschild solution}\label{secISS}

In order to model the gravitational collapse of \textit{dust} in the Quantum Einstein Gravity approach to asymptotic safety we will assume the existence of a spherically symmetric spacetime $\mathcal V$ in which the collapse takes place. We will also assume that the spacetime will be split into two different regions $\mathcal V =\mathcal V^+ \cup \mathcal V^-$ with a common spherically symmetric time-like boundary $\Sigma= \partial \mathcal V^+ \cap \partial \mathcal V^-$, corresponding to the surface of the star.
With regard to the stellar exterior region $\mathcal V^+$, we will describe it with a portion of an \textit{improved Schwarzschild solution}. Specifically, we are choosing for the exterior region an effective improved solution coming from the asymptotic safety approach that incorporates quantum corrections to the classical solution
(\cite{B&RIS} and references therein). A summary of this effective solution could be the following. The spacetime metric for this solution can be written as
\begin{equation}\label{RGISch}
ds_+^2=-\left(1-\frac{2 G(\bar R) M}{\bar R}\right) dt_S^2+\left(1-\frac{2 G(\bar R) M}{\bar R}\right)^{-1} d\bar R^2+ \bar R^2 d\Omega^2.
\end{equation}
where
\begin{equation}
G(\bar R)=\frac{G_0 \bar R^3}{\bar R^3+\tilde{\omega} G_0 (\bar R+\gamma G_0 M)},\label{runningG}
\end{equation}
$G_0$ is Newton's universal gravitational constant, $M$ is the mass measured by an observer at infinity and $\tilde{\omega}$ and $\gamma$ are constants coming from the non-perturbative renormalization group theory and from an appropriate cutoff identification, respectively.
The qualitative properties of this solution are fairly insensitive to the precise value of $\gamma$. However, in \cite{B&RIS}\cite{B&RIV} it is argued that the preferred value for $\gamma$ is $\gamma=9/2$.
On the other hand, $\tilde \omega$ can be found by comparison with the standard perturbative quantization of Einstein's gravity (see \cite{Dono} and references therein). It can be deduced that its precise value is $\tilde \omega=167/30\pi$, but again the properties of the solution do not rely on its precise value as long as it is strictly positive.

If we define
\[
\chi\equiv 1-\frac{2 G(\bar R) M}{\bar R},
\]
the horizons of the improved solution can be found by solving $\chi=0$.
It is easy to see that the horizons correspond to the number of positive real solutions of a cubic equation and depend on the sign of its discriminant or, equivalently, on whether the mass is bigger, equal or smaller than a critical value
\[
M_{cr}=\frac{1}{24} \sqrt{\frac{1}{2} (2819+85 \sqrt{1105})}\sqrt{\frac{\tilde\omega}{G_0}}\simeq 2.21 \sqrt{\tilde{\omega}} m_p \simeq 2.94 m_p,
\]
where $m_p$ is Planck's mass.
If $M>M_{cr}$ then the equation $\chi=0$ has two positive real solutions $\{\bar R_I,\bar R_O\}$ satisfying $\bar R_I<\bar R_O$.
The existence of an \emph{inner} solution $\bar R_I$ represents a novelty with regard to the classical spacetime.
However, it is interesting to remark that it is a result common to different approaches to Quantum Gravity. (See, for example, \cite{Modes}\cite{Amel}\cite{Nicol}).
The \emph{outer} solution $\bar R_O$ can be considered as the \textit{improved Schwarzschild horizon}, i.e., the Schwarzschild horizon with quantum corrections taken into account. The `improvement' in this horizon is, however, negligible for stellar masses, as can be made apparent
if one expands $\bar R_O$ in terms of $m_p/M$ obtaining
\[
\bar R_O\simeq 2 G_0 M \left[1-\frac{(2+\gamma)}{8} \tilde\omega\ \left(\frac{m_p}{M}\right)^2\right].
\]

In order to interpret the physical meaning of this solution let us suppose that it has been generated by an effective matter fluid in such a way that the coupled gravity-matter system satisfies Einstein's equations $G_{\mu\nu}=8\pi G_0 T_{\mu\nu}$ \cite{B&RIS}\cite{BHInt}.
Consider now a radially moving observer with an arbitrary 4-velocity $\bar{\mathbf{u}}$ and an orthonormal basis
$\{ \bar{\mathbf{u}},\bar{\mathbf{n}}, \mbox{\boldmath{$\omega$}}_\theta, \mbox{\boldmath{$\omega$}}_\varphi \}$ such that
\mbox{\boldmath{$\omega$}}$_\theta \equiv \bar R^{-1} \ \partial/\partial\theta$,
\mbox{\boldmath{$\omega$}}$_{\varphi}\equiv (\bar R\sin\theta)^{-1} \ \partial/\partial\varphi$ and $\bar{\bf n}$ is a space-like 4-vector.
The radially moving observer will write the vacuum energy-momentum tensor as an anisotropic fluid
\begin{eqnarray}\label{TV}
\mathbf{T}^+ =& \varrho_V \bar{\mathbf{u}} \otimes
\bar{\mathbf{u}} + p_V \bar{\mathbf{n}}\otimes \bar{\mathbf{n}}+ p_{\bot}
(\mbox{\boldmath{$\omega$}}_\theta \otimes \mbox{\boldmath{$\omega$}}_\theta +
\mbox{\boldmath{$\omega$}}_\varphi \otimes \mbox{\boldmath{$\omega$}}_\varphi)\ , \label{VIV}
\end{eqnarray}
where $\varrho_V$ is the vacuum energy density, $p_V$ is the vacuum normal pressure and $p_{\bot }$ is the vacuum tangential pressure.
By using the field equations, one can obtain their explicit expressions:
\begin{eqnarray}
\varrho_V &=& \frac{M G,_{\bar R}}{4 \pi G_0 \bar R^2} =-p_V, \label{varrho}\\
p_{\bot } &=& -\frac{M G,_{\bar R\bar R}}{8 \pi G_0 \bar R},\nonumber
\end{eqnarray}
where $G,_{\bar R}$ and $G,_{\bar R\bar R}$ are, respectively, the first and second derivatives of $G$ with respect to $\bar R$.

\section{Improved dust interiors}\label{secInterior}

In order to obtain the complete stellar model, we are now searching for the general class of quantum improved interiors $\mathcal V^-$ made of non-interacting particles which are matchable with the improved exterior solution. In other words, $\mathcal V^+$and $\mathcal V^-$ should satisfy Darmois matching conditions on $\Sigma$, what implies that the interiors must be such that the first and second fundamental forms of $\Sigma$ must coincide when computed from $\mathcal V^+$ or $\mathcal V^-$ \cite{Darmois}\cite{FST}.

Locally, every spherically symmetric spacetime metric can be written in geodesic coordinates as
\begin{equation}\label{genemetric}
ds_-^2=-d\tau^2+f(\tau,r) dr^2+ R(\tau,r)^2 d\Omega^2,
\end{equation}
where, if the spacetime is filled with a fluid, $\tau$ is the proper time of the particles composing the fluid and $r$ is a parameter that labels every shell of the fluid.

The matching of the interior solution to the improved Schwarzschild exterior will be performed through a spherically symmetric time-like hypersurface $\Sigma$ comoving with the fluid. I.e., the stellar surface will be defined by choosing a matching shell $r=r_\Sigma$. Since we do not have energy entering or leaving the star, the total mass $M$ of the star in the matched model should be completely determined by the value chosen for $r_\Sigma$, i.e., $M=M(r_\Sigma)$.

Darmois matching conditions and, in particular,
the requirement that the first fundamental forms of $\Sigma$ must coincide
implies that the \textit{areal radii} for the interior ($R$) and exterior regions ($\bar R$) must agree on $\Sigma$ \cite{FST}:
\begin{equation}\label{Rs}
R(\tau,r)\stackrel{\Sigma}{=}\bar R.
\end{equation}

On the other hand, another consequence of the matching conditions is that the \textit{mass functions} \cite{M&S,mass,Hayward} at both sides of the matching hypersurface $\Sigma$ must coincide \cite{FST}. The mass function of the interior solution is defined by $\mathcal M^- \equiv R (1-g_-^{\alpha\beta} \partial_\alpha R \partial_\beta R)/(2 G_0)$, what allows us to write $f$ (for later use) as
\begin{equation}\label{fmu}
f=\frac{R'^2}{\dot{R}^2+1-2 G_0 \mathcal M^-/R},
\end{equation}
where the apostrophe in $R'$ denotes derivative with respect to $r$ and the overdot in $\dot{R}$ denotes derivative with respect to $\tau$.
For the exterior, the mass function takes the form $\mathcal M^+ \equiv  \bar R (1-g_+^{\alpha\beta} \partial_\alpha \bar R \partial_\beta \bar R)/(2 G_0)= M(r_\Sigma) G(\bar R)/G_0$, so that
we will have on the matching surface
\begin{equation}\label{massS}
\mathcal M^-(\tau,r_\Sigma) \stackrel{\Sigma}{=} M(r_\Sigma) G(\bar R)/G_0.
\end{equation}

\begin{figure}
\includegraphics[scale=1.2]{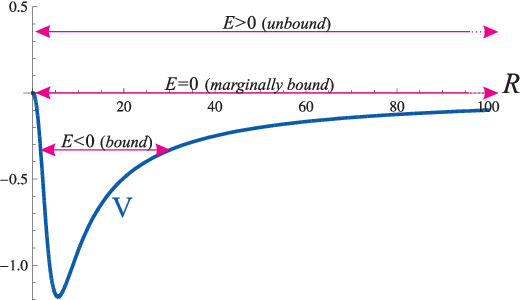}
\caption{\label{figBondi} For a given shell (i.e., a particular $r$) we have plotted a generic function $V(R)\equiv -G(R) M(r)/R$. From (\ref{Energies}) one deduces that in the $E>0$ case the behaviour of the shell would be unbounded. It would marginally bounded for $E=0$, since an expanding shell would reach infinity with $\dot R=0$. Finally, in the $E<0$ case the shell is \textit{doubly} bounded by a minimum and a maximum radius. Note that, as in the classical case, there is a limiting lower value for $E$ given by $E_{lim}=-1/2$ and coming from the signature in (\ref{ImprovedLTB}) and not by the minimum of $V$.}
\end{figure}

One of the features of dust interiors in GR that differentiate them from interiors possessing general fluids is that, provided one has a dust solution, one can match it with an exterior Schwarzschild solution (with non-fixed mass\footnote{The mass of the model (and, therefore, of the Schwarzschild solution) is fixed once one chooses a particular $r_\Sigma$.}) at an arbitrary value $r_\Sigma$. This is so because the matching conditions require the normal pressures to coincide on $\Sigma$ when computed from $\mathcal{V}^+$ or $\mathcal{V}^-$ \cite{Israel}\cite{FST}, what is trivially satisfied in a classical dust model in which the pressures are zero everywhere. Of course, in the quantum improved case, the normal pressures must also coincide on $\Sigma$. However, now there is a non-zero exterior vacuum pressure and, consequently, there must be a non-zero interior pressure which, by definition of dust, we do not want to be attributable to the non-interacting particles, but only to quantum effects.
Interestingly, although not surprisingly, this can be satisfied if one assumes that the matching between an improved dust interior and the improved Schwarzschild exterior can be performed at an arbitrary $r_\Sigma$, as in the classical case.
In order to check this statement note that, in view of (\ref{massS}) and (\ref{Rs}), this requirement means that the inner mass function should take the form
\begin{equation}\label{mu-}
\mathcal M^-(\tau,r) = M(r) G/G_0,
\end{equation}
where
\begin{equation}
G=G(\tau,r)=\frac{G_0 R^3}{R^3+\tilde\omega G_0 (R+\gamma G_0 M(r))}\label{Gdust}.
\end{equation}
If we denote the 4-velocity at the star surface as $\mathbf u\ (=\partial/\partial\tau)$ and the normal vector to $\Sigma$ as $\mathbf{n}\ (=f^{-1/2} \partial/\partial r)$ the matching conditions \cite{Israel}\cite{FST} require that $T_{\alpha\beta} n^\alpha u^\beta$ should coincide when computed from the interior or the exterior. In the case of the exterior (\ref{TV}) $T^+_{\alpha\beta} {n^+}^\alpha {u^+}^\beta = 0$, what means that there is not a flux of energy in the normal direction, as expected. This requires that  $T^-_{\alpha\beta} {n^-}^\alpha {u^-}^\beta = 0$, what, with the help of (\ref{fmu}) and (\ref{mu-}) provide us with the condition
\begin{equation}\label{Newton}
\ddot{R} = - \frac{G M(r)}{R^2} + \frac{M(r) G,_R}{R}.
\end{equation}
Note that the second term of this equation admits the interpretation of the sum of a gravitational (attractive) term plus a quantum (repulsive, since (\ref{runningG}) implies $G,_R\geq 0$) term.
The differential equation (\ref{Newton}) can be integrated providing us with
\begin{equation}\label{Energies}
\frac{\dot{R}^2}{2}=\frac{G M(r)}{R}+ E(r),
\end{equation}
where $E(r)$ is an arbitrary function coming from the integration. The form of (\ref{Energies}) indicates that it admits an interpretation \`{a} la Bondi \cite{Bondi}: By choosing a function $E(r)$ one chooses the total energy per unit mass of the particles in the fluid within a shell of radius $r$. Then, (see figure \ref{figBondi}) if $E > 0$ the system would be \emph{unbound}, if $E = 0$ the system would be \emph{marginally bound} and if $E < 0$ the system would be \emph{bound}. In fact, the only novelty with respect to the usual LTB solutions is that now the $E<0$ is \textit{doubly bound}, meaning that there will not only be an upper bound to the particles, but also a non-zero lower bound where the shell will bounce\footnote{In this doubly bound case ($-1/2<E<0$), for every shell $r_i$, this corresponds to the minimum value $R_B$ such that $G(R_B) M(r_i)/R_B+E(r_i)=0$, that will always exist since $\lim_{R\rightarrow 0} G(R) M(r)/R=0$, as is clear in fig.\ref{figBondi}.}.

Finally, by using (\ref{fmu}) and the help of (\ref{Energies}) we can particularize the interior spacetime metric (\ref{genemetric}) for our dust interiors arriving at the following
\begin{prop}
The class of `improved dust interiors' is defined by the metric
\begin{equation}\label{ImprovedLTB}
ds_-^2=-d\tau^2+\frac{R'(\tau,r)^2}{1+2 E(r)} dr^2+ R(\tau,r)^2 d\Omega^2.
\end{equation}
with $R(\tau,r)$ determined by (\ref{Energies}).
\end{prop}

In order to evaluate the physical content of this solution we proceed as with the exterior solution (sect. \ref{secISS}), but fixing $\mathbf{u}$ to be the 4-velocity of the fluid particles ($\mathbf{u}=\partial/\partial\tau$) and fixing $\mathbf{n}$ accordingly. In this way one obtains
\begin{eqnarray}\label{TEMint}
\mathbf{T}^- =& \rho\ \mathbf{u} \otimes
\mathbf{u} + p\ \mathbf{n}\otimes \mathbf{n}+ p_{\bot}
(\mbox{\boldmath{$\omega$}}_\theta \otimes \mbox{\boldmath{$\omega$}}_\theta +
\mbox{\boldmath{$\omega$}}_\varphi \otimes \mbox{\boldmath{$\omega$}}_\varphi)\ ,
\end{eqnarray}
where
\begin{eqnarray}
\rho &=& \frac{G M'}{4 \pi G_0 R^2 R'}+\frac{ M G'}{4 \pi G_0 R^2 R'}\label{rho}\\
p &=& -\frac{ M G,_R}{4 \pi G_0 R^2}\label{pr}\\
p_{\bot } &=& -\frac{(M G,_{R})'}{8 \pi G_0 R R'}.\label{ptan}
\end{eqnarray}
\begin{rem}
The pressures in the improved dust interiors are only attributable to the effect of the quantum correction.
\end{rem}
In other words, as expected for a dust solution, if we turn off the quantum improvements (i.e., if $G=G_0=$constant) the only non-zero term left would be the usual dust density in the LTB solutions (first term in the r.h.s. of (\ref{rho})).
On the contrary, if the mass becomes constant we would have eliminated the stellar interior and we would be left with the improved black hole vacuum densities and pressures (\ref{TV}).
\begin{rem}
As required by the matching conditions \cite{Israel}\cite{FST}, the normal pressures in the matched model coincide when computed from $\mathcal V^+$ (\ref{varrho}) or $\mathcal V^-$ (\ref{pr}).
\end{rem}

\section{Absence of shell-focusing singularities}\label{secAbsSing}

The classical LTB solutions possess two different types of singularities. On the one hand, those in which $R'=0$, what means that the radial geodesic distance between two different shells becomes zero at a certain instant and which are consequently called \textit{shell-crossing singularities}. These singularities have been analyzed in detail in the literature \cite{H&L}\cite{MY&S} and have been shown to be gravitationally weak \cite{Newman}. In fact, they can be avoided with an appropriate choice of the arbitrary functions in the model. In this way, they are generally believed to be a mathematical artifact of the (pressureless) model and we will not consider them further here. On the other hand, there are the singularities in which $R=0$ that correspond to the usual image of a `central singularity' generated by gravitational collapse and which are called \textit{shell-focusing singularities}. In the classical LTB solutions they can be gravitationally strong and hidden/naked to different observers. In what follows it will be our goal to show the absence of these kind of singularities in the improved dust models.

\begin{lemma}
If $R'\neq 0$, the energy-momentum tensor of the improved dust solutions is bounded. In particular, it is bounded when the shells approach $R=0$.
\end{lemma}

In the absence of shell-crossing, the densities and pressures of the improved dust model can only be unbounded as the shells approach $R=0$, but since, around $R=0$, $G(R)\propto R^3$ it is easy to check using (\ref{rho}), (\ref{pr}) and (\ref{ptan}) that they all take a finite Planckian value as the shells collapse towards the center:
\[
\rho \simeq -p\simeq-p_{\bot }\simeq \frac{3}{4 \pi G_0^2 \gamma \tilde\omega },
\]
what implies that the solutions are devoid of shell-focusing matter singularities.

\begin{lemma}
If $R'\neq 0$, there are not scalar curvature singularities in the improved dust solutions as the shells approach $R=0$.
\end{lemma}

A spacetime does not possess scalar curvature singularities if the scalar invariants polynomial in the Riemann tensor remain finite. A full independent set of invariants was found in \cite{Carmi}. For spherically symmetric spacetimes evaluated at $R=0$ one can take as the algebraically independent set of scalars $\mathcal{R}$ and $r_1$ \cite{Charact}, where $\mathcal{R}$ is the curvature scalar and  $r_1\equiv {S_\alpha}^\beta {S_\beta}^\alpha$, being ${S_\alpha}^\beta \equiv {R_\alpha}^\beta-{\delta_\alpha}^\beta \mathcal{R}/4$ the trace-free Ricci tensor and ${R_\alpha}^\beta$ the Ricci tensor.

Using the improved dust solution (\ref{ImprovedLTB}) and the help of (\ref{Gdust}), (\ref{Newton}) and  (\ref{Energies}) one gets the following behaviour for the algebraically independent scalars when the shells approach $R=0$
\[
\lim_{R\rightarrow 0} \mathcal{R}=\frac{24}{G_0 \gamma \tilde{\omega}} \hspace{1cm}\mbox{and}\hspace{1cm}
\lim_{R\rightarrow 0} r_1= 0,
\]
what shows that no scalar curvature singularities will be generated.
Finally, the above lemmas can be summarized in the following
\begin{prop}
The improved dust models do not possess shell-focusing singularities.
\end{prop}

\section{Conclusions}\label{secConc}

In this work, a class of solutions interpretable as composed of quantum corrected dust has been obtained (rem.1). This class possesses a running gravitational constant $G(R)$ (\ref{Energies}) compatible with the asymptotic safety scenario. In turn, these \textit{improved dust solutions} have inherited $G(R)$ from an \textit{improved Schwarzschild solution} that is fully matchable to them ((\ref{Rs}),(\ref{massS}) and rem.2 \cite{FST}). Since the running gravitational constant in the improved Schwarzschild solution comes from the antiscreening effect of virtual gravitons \cite{B&RIS}, it is natural to interpret (by comparing (\ref{TV}) with (\ref{TEMint})) that the new $G,_R$-and/or-$G,_{RR}$-dependent corrections to the matter content of the dust come directly from this antiscreening effect. This interpretation is also supported by the appearance of a typical repulsive force term in (\ref{Newton}) which acts against the gravitational collapse. In fact, this repulsive term is the ultimate responsible for the bounce of shells in the $E<0$ case. However, it cannot avoid the collapse in the $E\geq 0$ case. Notwithstanding this fact, we have also seen that the collapsing shells never generate a shell-focusing singularity. This has been shown by discarding both matter and scalar curvature singularities.

As mentioned in the introduction, a previous approach to this subject \cite{Casadio} suggested that the asymptotic safety approach could not be able to prevent the generation of singularities during the collapse. Specifically, the authors analyzed a dust model with $R= r a(\tau)$ and $M(r)=\mu r^3$ (being $\mu$ a constant) through a heuristic approach inspired by asymptotic safety (in which the running gravitational constant was assumed to depend only on the dust density) and they were able to obtain divergent dust densities.
The corresponding \textit{improved dust model},
however, offers a finite density (\ref{rho}) satisfying $\lim_{R\rightarrow 0} \rho = 3/(2 \pi G_0^2 \gamma \tilde{\omega})$ and is, in general, devoid of any kind of singularity.

The purpose of this work has been to show that the asymptotic safety approach might offer a solution to the problem of singularities in gravitational collapse. Of course, the models presented here have several simplifications (symmetry, non-interacting matter, absence of Hawking radiation, etc.) that should be surpassed in future works \cite{ToPa} in order to get more realistic stellar models. However, the fact that the results of this work have been obtained for the specially problematic case of dust
suggests that a more realistic model could hardly contradict the main result in this work: the absence of shell-focusing singularities.

\end{document}